\documentclass[twocolumn,showpacs,preprintnumbers,amsmath,amssymb]{revtex4}

\usepackage{color}
\usepackage{amsmath}
\usepackage{graphicx}
\usepackage{dcolumn}
\usepackage{bm}

\newcommand{\ew}[1]{\langle #1 \rangle}
\newcommand{\beq}{\begin{eqnarray}}
\newcommand{\eeq}{\end{eqnarray}}

\newcommand{\eq}[1]{Eq.~(\ref{#1})}

\begin{document}

\title{Nonequilibrium Quantum Phase Transitions in the Dicke Model} 
\author{V. M. Bastidas}
\author{C. Emary}
\author{B. Regler}
\author{T. Brandes}

\affiliation{%
Institut f\"ur Theoretische Physik, Technische Universit\"at Berlin, Hardenbergstraße 36, D-10623 Berlin, Germany}%

\date{\today}

\begin{abstract}

We establish a set of nonequilibrium quantum phase transitions in the Dicke model by considering a monochromatic nonadiabatic modulation of the atom-field coupling.
For weak driving the system exhibits a set of sidebands which allow the circumvention of the no-go theorem which otherwise forbids the occurence of superradiant phase transitions.  
At strong driving we show that the system exhibits a rich multistable structure and exhibits both first- and second-order nonequilibrium quantum phase transitions.
\end{abstract}


\maketitle

The Dicke model (DM) is a paradigm of collective behavior in quantum mechanics \cite{Dicke} and describes the interaction of $N$ two-level atoms with a single-mode bosonic field. The DM undergoes a quantum phase transition (QPT) at a critical atom-field coupling, the same order of magnitude as the atomic level splitting \cite{Hepp-Lieb, Brandes}. 
Whereas in cavity QED it
is forbidden by a ``no-go theorem''  
\cite{Ciuti,Marquardt}, the superradiant QPT has recently been realized experimentally in a simulation of the DM with a Bose-Einstein condensate in an optical cavity \cite{Esslinger}.    

Even under an adiabatic evolution many-body systems can exhibit highly nontrivial behavior \cite{Altland,Polkovnikov,Zurek}.
In particular, spontaneous symmetry breaking at the DM QPT in a Bose-Einstein condensate was observed experimentally by adiabatically crossing the critical point \cite{Mottl}. Furthermore, nonadiabatic evolution of quantum systems has attracted considerable interest, both theoretical and experimental\cite{Hangui, Holthaus, Arimondo,Gunter}.

In this Letter we study a driven version of the DM in which we assume a time-dependent atom-field coupling.  In comparison with previous works which consider nonadiabatic modulation of a single two-level system \cite{DeLiberato,CarmichaelAlsing,Peano,Grifoni} or an $N$-atom system under adiabatic modulation \cite{Altland,Vedral}, we address here the fundamental issue of the influence of nonadiabatic modulation on QPTs.

In the limit of weak driving strength we show that our driven DM exhibits a set of new nonequilibrium normal-superradiant QPTs (QPT sidebands) when the driving is near resonance with the excitation energies of the undriven system.  The QPTs are of second order and similar in kind to the original Hepp-Lieb transition \cite{Hepp-Lieb}.  
We show that these nonequilibrium QPTs are not forbidden by the  no-go theorem, thus bringing the otherwise-forbidden Dicke-type QPT in the realm of observability in cavity and circuit QED setups.
Our analysis also allows us to go beyond this perturbative regime and investigate the limit of strong driving.  In this regime we show that, in comparison with  previous proposals for driven QPTs \cite{Creffield,Lindner,Tanaka}, the nature of criticality changes dramatically. We find a rich nonequilibrium phase diagram replete with a host of macroscopically-distinct meta-stable phases and a nonequilibrium first-order QPT with no analogues in the static case.
\\

\textit{Driven Dicke model.---} 
Following Dicke \cite{Dicke}, we describe an ensemble of $N$ identical, distinguishable two-level atoms (level splitting $\omega_{0}$) by means of collective operators $J_{\alpha}; \alpha \in \{z,+,-\}$, which obey the angular momentum commutation relations (with cooperation number $j=N/2$).  These atoms interact with a bosonic mode of frequency ${\omega}$ via a dipole interaction.  With the atom-field coupling strength time-dependent, we obtain the {\em driven} DM:
\begin{eqnarray}
      \label{drivendicke}
            \hat{H}(t)&=&\omega
            a^{\dagger}a +\omega_{0}J_{z}+\frac{g(t)}{\sqrt{N}}(a^{\dagger}+a)\left (J_{+}+J_{-}\right)
      ,
\end{eqnarray}
and in the following we shall consider a monochromatic modulation with a static contribution: $ g(t)= g + \Delta g \cos \Omega t$.

\textit{Normal-phase stability.---} 
In the thermodynamic limit, $N \to \infty$, the undriven ($\Delta g =0$) DM exhibits a QPT at a critical coupling $g_c=\sqrt{\omega\omega_{0}}/2$, where the ground state changes from an unexcited {\em normal phase} to a symmetry-broken {\em superradiant phase} in which both the field and atomic collection acquire macroscopic occupations \cite{Hepp-Lieb, Brandes}.
We begin our analysis by investigating the stability of this normal phase under driving.
To this end, we construct a normal-phase effective Hamiltonian in the same way as in Ref.~\cite{Brandes} for the undriven case:  we make a Holstein-Primakoff representation of the angular momentum algebra in terms of bosonic operators $b,b^\dag$ and take the thermodynamic limit, assuming $b \sim N^0$.  The result is the driven normal-phase Hamiltonian
$
  \hat{H}_\mathrm{NP}(t) = \omega_0 b^\dag b + \omega a^\dag a + g(t)(a^\dag + a) (b^\dag + b )
$, which describes fluctuations about the vacuum cavity state and unexcited atomic ensemble.
In the Heisenberg picture, the  equations of motion for the normal coordinate operators of this model, $\hat{q}_{\pm}(t)$, read \cite{Bastidas10}
\begin{eqnarray}
      \label{eqn:Floquet_Mathieu}   
	 \ddot{\hat{q}}_{\pm}(t) + \biggl [\varepsilon_{\pm}^2 \pm 2 \omega \Delta g \cos \Omega t\biggr ] \hat{q}_{\pm}(t) = 0
      , 
\end{eqnarray}
with $\varepsilon_{\pm} = \sqrt{\omega^2 \pm 2 g \omega } $, the excitation energies of the undriven normal phase and where we have set $\omega_{0} = \omega$ for simplicity.
Equation (\ref{eqn:Floquet_Mathieu}) represents two uncoupled Mathieu equations \cite{Arnold}. 
In classical dynamics, the Mathieu equation exhibits the phenomenon of parametric resonance and has stable and unstable solutions whose location is given by the Arnold tongues \cite{Arnold}. The manifestation of parametric resonance in the quantum regime has also been studied e.g., \cite{Popov}.

In our case, when both normal modes $\hat{q}_{\pm}$ are stable, Hamiltonian $H_\mathrm{NP}(t)$ permits bound solutions, localized around unoccupied field and atomic modes.  Either scaled field occupation, $2\ew{a^{\dagger} a}/N$, or scaled atomic occupation, $2\ew{b^{\dagger} b}/N= 2\ew{J_z}/N +1$,  may be taken as the order parameter for this system, and here they are both zero ($N \to \infty$).  When unstable,  $H_\mathrm{NP}(t)$  possesses only unbounded solutions and ceases to be a valid approximation to the full Hamiltonian (admitting the possibility that the order-parameter becomes finite).
%

\begin{figure}
\centering
\includegraphics[width=0.48\textwidth]{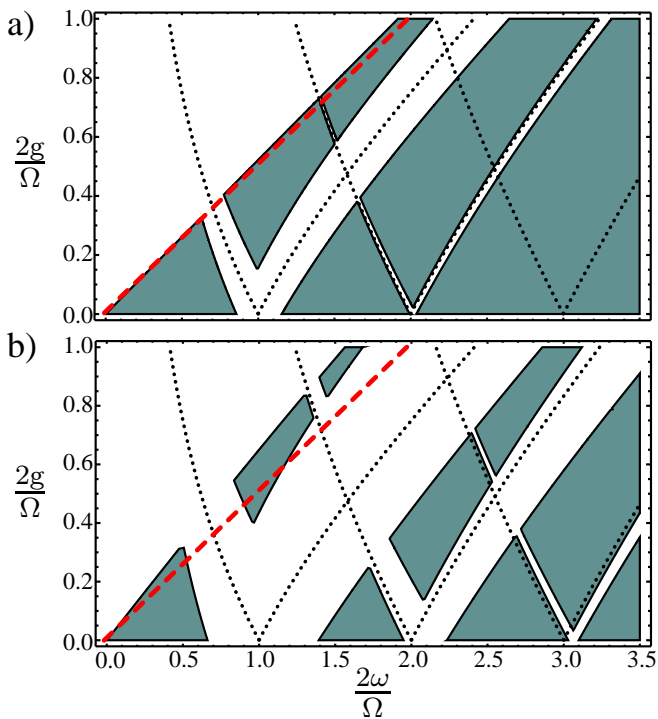}
\caption{
  \label{fig_NP}
  Stability diagram of the nonequilibrium normal phase on resonance $\omega=\omega_0$  for  (a) weak driving, $\Delta g/\Omega=0.15$, and (b) strong driving, $\Delta g/\Omega=0.4$.
  The colored zones correspond to stability, white zones, instability. 
  In the undriven case, the separatrix between stable and unstable zones is given by $g_c=\omega/2$ on resonance (dotted red line).
  For weak driving instability zones open up around the resonance conditions $\varepsilon_{\pm} = k (\Omega/2) $ for $k=1,2,3$ (dotted black lines), which grow and begin to dominate for large driving.
}
\end{figure}

Figure \ref{fig_NP} depicts these stable zones as a function of the static parameters $g$ and $\omega$ --- the colored zones correspond to stability, white zones, instability.  
Without driving, $\Delta g = 0$, the stability-instability transition corresponds to the standard DM QPT along the line $g=\omega/2$ (on resonance).  Increasing the driving strength $\Delta g$ has two effects: it leads to a shift of this critical line---as a consequence of parametric stabilization \cite{Arnold}---and more importantly, it opens up new zones of instability in the normal phase. 
The precise locations of these zones can be obtained from the known behavior of the Arnold tongues \cite{Arnold}. 
When they first appear, these zones are located around resonance between driving frequency and (undriven) excitation  energies: $k \Omega =2 \varepsilon_{\pm}$ with integer $k \geq 0$.
A similar instability for $k=1$ was briefly discussed in the dispersive limit $\omega \gg \omega_0$ in Ref. \cite{Tureci}. For sufficiently small driving, the width of the $k$th unstable zone scales like $(2/\Omega)^{2k-1}(\Delta g\ \omega)^k$.  For large $\Delta g$ the new instability zones dominate the parameter space [Fig.~\ref{fig_NP}(b)].
Just as the change in stability of the undriven normal-phase effective Hamiltonian indicates the occurrence of a QPT, we interpret the change in the stability of $H_\mathrm{NP}(t)$ as ushering the occurrence of a nonequilibrium QPT.

\textit{Nonequilibrium QPT sidebands.---} 
To obtain more information about the nonequilibrium QPTs and the unstable zones of Fig.~\ref{fig_NP}, we employ Floquet theory \cite{Shirley}
and make use of a generalized rotating wave approximation, similar to that found in Refs.~\cite{Irish, Nori, Grifoni}. 
Motivated by the fact that for small static coupling $g$, the $k$th instability zone arises close to $\omega \approx k\Omega/2$, we perform a canonical transformation of Eq. \eqref{drivendicke},
$
  \hat{H}(t) \to \hat{H}_{k}(t) = \hat{U}_{k}^\dagger(t)[\hat{H}(t)-i \frac{\partial}{\partial t}]\hat{U}_{k}(t)
$
with
\begin{equation}
      \label{UnitaryRot}
            \hat{U}_{k}(t) =
            e^{-i\mu(t)(a^{\dagger}+a)J_{x}}  
            e^{-i\frac{k\Omega}{2} (J_z +a^\dagger a) t}
      ,
\end{equation}
where $\mu(t)=\frac{2}{\sqrt{N}}(\frac{\Delta g}{\Omega})\sin\Omega t$.
The explicit calculation of $\hat{H}_{k}(t)$ is reported in Ref.~\cite{SI}.  Here it suffices to note that it can be written in the form
\begin{equation}
   \hat{H}_{\text{k}}(t)=\sum_{n=-\infty}^{\infty} \hat{h}^{(k)}_{n}\exp{(i n \Omega t)}  \ .
\end{equation}
In analogy with the standard RWA of quantum optics, we obtain an approximate Hamiltonian to describe the $k$th resonance by neglecting all terms in $\hat{H}_{k}(t)$ with oscillatory time-dependence:  $\hat{H}_{k}(t) \approx  h_{0}^{(k)} $.
We therefore approximate our time-dependent problem with a series of time-independent models, one for each $k\ge 0$.  The eigenenergies of these time-independent Hamiltonians are actually {\em  quasienergies} \cite{Shirley}.

We first consider the fundamental $k=0$ case with approximate Hamiltonian
\begin{eqnarray}
      \label{ZeroResonaceHam}
             h^{(0)}_{0} &=& \delta^{(0)}
            a^{\dagger}a +\delta^{(0)}_{0}\mathcal{J}_{0}\left(\frac{2\Delta g}{\Omega\sqrt{N}}(a^{\dagger}+a)\right)J_{z}
            \nonumber \\&& 
            +\frac{g}{\sqrt{N}}(a^{\dagger}+a)\left(J_{+}+J_{-}\right)
            + 2\omega\left (\frac{\Delta g}{\Omega}\right)^{2}\frac{J_{x}^{2}}{N}
      ,
\end{eqnarray}
where $\delta^{(k)}=\omega-k\Omega/2$ and $\delta^{(k)}_{0}=\omega_{0}-k\Omega/2$ are detunings in the $k$th rotating frame, and $\mathcal{J}_0(x)$ is the zeroth-order Bessel function, here, as in Ref.~\cite{Hangui}, with operator argument. The validity of the approximation $\hat{H}_{0}(t) \approx  h_{0}^{(0)} $ requires $g,\ \omega ,\ \omega_0 \ll \Omega,\Omega\max(1,\Delta g/\Omega)$\cite{SI}.
In the limit $\Delta g\rightarrow 0$, this Hamiltonian recovers the original undriven Dicke model.  Expanding up to second order in $\Delta g/\Omega$, we obtain a model similar to the undriven DM \cite{SI}, that can be analyzed in the same way \cite{Brandes}. This analysis reveals a second-order QPT along the critical line $g=\omega/2+\omega(\Delta g/ \Omega)^2$ ($\omega=\omega_0$), which gives the first contribution to the shift of the DM phase boundary observed in Fig.\ref{fig_NP}.
For the higher resonances, $k>0$, exact expressions for $h^{(k)}_{0}$ are difficult to obtain.  However, it is possible to write down explicit expansions up to any finite order in $\Delta g/\Omega$.  To second order, the $k=1$ Hamiltonian reads
\begin{eqnarray}
      \label{FirstResonaceHam}
            h^{(1)}_{0} &=& \delta^{(1)}
            a^{\dagger}a +\delta^{(1)}_{0}J_{z}
            \nonumber \\&& \nonumber  +\frac{g}{\sqrt{N}}\left(a^{\dagger}J_{-}+aJ_{+}\right)+\frac{\Delta g}{2\sqrt{N}}\left (a^{\dagger}J_{+}+aJ_{-}\right)
            \nonumber \\&& 
            -2\omega_{0}\left(\frac{\Delta g}{\Omega}\right)^2 a^{\dagger}a \frac{J_{z}}{N}+\omega\left(\frac{\Delta g}{\Omega} \right)^2 \frac{J_{-}J_{+}}{N} 
      .
\end{eqnarray}
The region of validity of this Hamiltonian is $\omega \approx \omega_0$ and $\delta^{(1)},\ \delta^{(1)}_{0},\ g,\ (\Delta g/\Omega)^{2} \ll \omega,\omega_0$.
The first two lines of this Hamiltonian represent a model similar to the original Dicke model in which energy conserving and nonconserving parts of the interaction have independent coupling parameters ($g$ and $\Delta g/2$, respectively) similar to \cite{AparicioSimons}. At second order in $\Delta g/\Omega$, new effective interactions arise:  the terms on the third line of \eq{FirstResonaceHam} can be interpreted as an effective dispersive atom-field interaction and a dipole-dipole interaction between the atoms. Analytically, this Hamiltonian may again be treated in the same way as the undriven DM, from which we observe a second-order superradiant transition occurring at the critical lines $g=-\Delta g/2+|\delta^{(1)}+\omega(\Delta g/ \Omega)^2|$.
A similar analysis for the $k=2$ case yields a further second-order QPT along the lines at $g=\delta^{(2)}+\frac{\omega}{2}(\Delta g/ \Omega)^2$ and $g=-\delta^{(2)}-\frac{3\omega}{2}(\Delta g/ \Omega)^2$.  This analysis can be repeated for the all values of $k$.

For small driving, thus, Fig.~\ref{fig_NP} shows that the original DM QPT is joined by a set of new nonequilibrium {\em QPT sidebands}, the visible number of which (i.e. have significant width) increases with increasing driving strength.  
Each of these nonequilibrium QPTs is similar to the original transition in many respects; the transitions are of the second-order, mean-field type, with the same critical exponents as in undriven DM \cite{Brandes}.
An important difference, however, is that, whereas the original DM QPT occurs for a static coupling $g\sim \omega/2$, the sideband QPTs occur for a coupling $g \sim \delta^{(k)}\ll \omega,\omega_0$  with driving strength also of the same magnitude. In contrast to the static case, since the detuning $\delta^{(k)}$ can be made arbitrarily small, the term $\hat{H}_{NG}=\alpha\frac{(g(t))^2}{\omega_0}(\hat{a}^{\dagger}+\hat{a})^2$ arising from the square of the vector potential can be neglected even for $\alpha>1$. 
Therefore, the sideband QPTs are not prohibited by the no-go theorem  \cite{Ciuti,Marquardt} and should thus be observable in e.g. cavity- and circuit- QED experiments \cite{Footnote1}. 

\begin{figure}
\centering
\includegraphics[width=0.48 \textwidth]{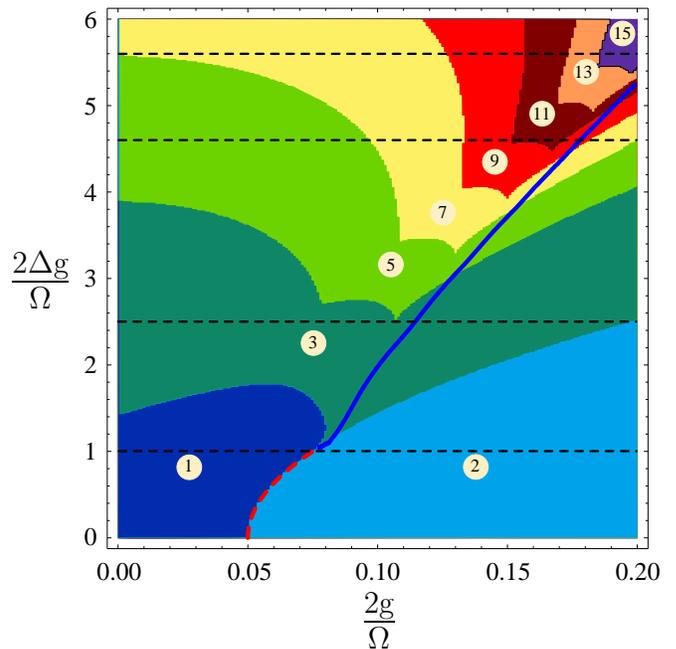}
\caption{
  \label{fig_multiphase}
  Phase diagram as a function of static coupling ($g$) and driving amplitude ($\Delta g$) of effective Hamiltonian $h^{(0)}_0$, which describes the driven DM near the $k=0$ resonance. The labels indicate the number of local minima of the ground-state energy landscape $E_G$ in the corresponding zone.
  The normal phase is the region where there is just a single minimum (
  $X=Y=0$).  The superradiant phase is a region with two global minima (nonzero order parameter).  The boundary between these two regions marks a second-order QPT (dotted red curve).  Outside these regions, the energy surface has an odd number $\ge 3$ of total minima, sometimes with a single global minima, sometimes with two.  The boundary between these possibilties is a first-order QPT (solid blue line). 
  The parameters are $\omega/\Omega = \omega_0/\Omega = 0.05$.
  }
\end{figure}

{\em First-order nonequilibrium QPT:-}
We now revisit the full (nonperturbative in $\Delta g/\Omega$)  $k=0$ resonance Hamiltonian, Eq.~\eqref{ZeroResonaceHam}.
We can investigate the critical properties of this model by using the Holstein-Primakoff representation again and by introducing macroscopic displacements $\sqrt{N/2}\ X$ and$\sqrt{N/2}\ Y$ of the field and atomic modes, respectively.
In the thermodynamic limit, $N \rightarrow \infty$, we obtain a quadratic effective Hamiltonian with  leading term $ \frac{1}{2} N E_G(X,Y)$ with
\begin{eqnarray}
  \label{LowestQuasienergy}
	E_G(X,Y) 
  &=&\!
  \omega \!\left(\frac{\Delta g}{\Omega}\right)^2 
    \!\!\! Y^2(2-Y^2) 
  - \frac{4g}{\sqrt{2}}XY\sqrt{2-Y^2} 
  \nonumber \\
  &&
  +\omega X^2   
  +  \omega_0 (Y^2-1) \mathcal{J}_{0}\left(\frac{4\Delta g X}{\sqrt{2}\Omega}\right)
  .
\end{eqnarray}
The global minima of $E_G(X,Y)$ give the ground-state energy of $h^{(0)}_{0}$ and either $X$ or $Y$ may be taken as order-parameter.  In the undriven case, the energy surface $E_G(X,Y)$ exhibits a bifurcation from single to double minima at the QPT.  In the driven case, while retaining this bifurcation, $E_G(X,Y)$ also develops multiple additional minima, with the general trend that the number of minima increases with driving strength (see Fig.~\ref{fig_multiphase}).

\begin{figure}
\centering
\includegraphics[width=1.03\linewidth]{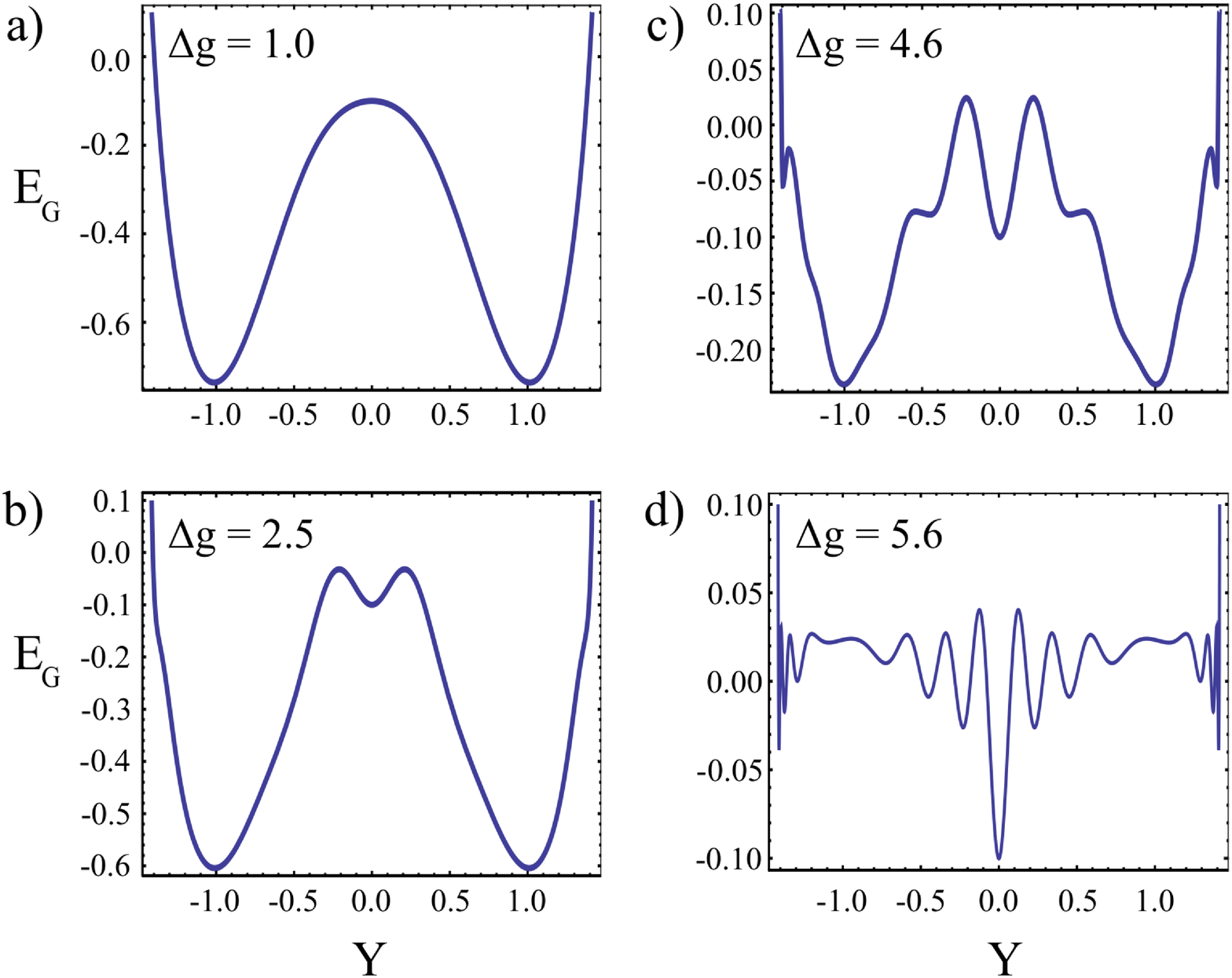}
\caption{
  \label{fig_potlines} 
  Sections of the ground-state energy landscape $E_G(X,Y)$ for a fixed value of the static coupling ($2g/\Omega = 0.195$) and increasing driving amplitudes ($\Delta g$) represented by dashed straight lines in Fig.~\ref{fig_multiphase}. 
  This sections correspond to the function $E_G(X[Y],Y)$, where $X[Y]$ is a line in the order-parameter space crossing all the critical points of the ground-state energy. 
  The four panels show
  (a) the superradiant phase with two global minima, 
  (b) the emergence of a local minimum at the origin,
  (c) multiple minima, but still two global minima, and
  (d) a single global minima at the origin plus many further local minima.
  Panels (c) and (d) are separated by the first-order phase transition.
  The parameters are $\omega/\Omega = \omega_0/\Omega = 0.05$.
}
\end{figure}

Figure~\ref{fig_potlines} shows a sequence of cuts through the energy surface for increasing driving strength with static coupling $g > g_c$.
As is clear from this sequence, at a certain value of $\Delta g$, the potential switches from having two global minima to just a single one at the origin.  At the boundary between these behaviors, the potential exhibits three global minima where normal-phase-like (with $X = Y = 0$) and superradiant-like (with $|X|,|Y| > 0 $) solutions coexist.  The system is therefore seen to exhibit a {\em first-order QPT} \cite{1OPT} as a function of driving strength, the location of which is shown in Fig.~\ref{fig_multiphase}.
Furthermore, we interpret the local minima of $E(X,Y)$ as metastable phases of $h^{(0)}_{0}$.  
These metastable states are related to the phenomenon of parametric stabilization \cite{Arnold}. 
Since these are separated from the global minima by macroscopic displacements, we expect transitions to the actual ground-state to be suppressed, such that the corresponding values of the order-parameters are observable.  This possibility is reinforced when one recalls that \eq{LowestQuasienergy} actually describes the lowest {\em quasienergy}, which does not have the same thermodynamic significance as the lowest actual energy.

In summary, then, we have discussed a driven Dicke model through the use of a series of 
effective Hamiltonians obtained under a generalized RWA.  For weak driving the system exhibits a set of QPT side-bands for which the no-go theorem is circumvented.  At strong driving the long-time dynamics of the system in the $k=0$ resonance case is governed by the effective Hamiltonian $h^{(0)}_{0}$, which exhibits rich structure with a first-order quantum phase transition and metastable states. We anticipate that the higher resonances show similar behavior.
Our methodology should be generalisable to investigate similar regimes for other phase transitions.

A possible experimental realization of our model could be carried out by means of a quantum well waveguide structure, which allows one to switch on the light-matter interaction within less than one cycle of light \cite{Gunter}. Another possible implementation consist in a Bose-Einstein condensate coupled to an optical cavity where the effective atom-field coupling could be controlled externally by varying the intensity of the pump laser as a function of time about a static value.

The authors gratefully acknowledge discussions with J. H. Reina, F. Brennecke, T. Esslinger, M. Hayn, V. A. Leyton, C. Nietner, V. Peano, G. Schaller, M. Thorwart, M. Vogel, and financial support from the DAAD and DFG Grants BR $1528/7-1$, $1528/8-1$, SFB $910$,
and GRK $1558$ .

\begin{widetext}

\section*{\large Supplementary Material for ``Non-equilibrium Quantum Phase Transitions in the Dicke Model''}

\subsection{The Hamiltonian in the rotating frame}
We study the general single-mode {\em driven} Dicke model:
\begin{eqnarray}
      \label{drivendickeSI}
            \hat{H}(t)&=&\omega
            a^{\dagger}a +\omega_{0}J_{z}+\frac{g(t)}{\sqrt{N}}(a^{\dagger}+a)\left (J_{+}+J_{-}\right) ,
\end{eqnarray}
and the case of a monochromatic modulation of the atom-field coupling $ g(t)=g+\Delta g \cos \Omega t$.
We  perform a canonical transformation of the Hamiltonian Eq. \eqref{drivendickeSI} into a rotating frame using
\begin{equation}
      \label{UnitaryRotSI}
            \hat{U}_{k}(t) =\exp\left[-i\frac{2}{\sqrt{N}}\left(\frac{\Delta g \sin\Omega t}{\Omega}\right)(a^{\dagger}+a)J_{x} \right]  \exp\left[-i\frac{k\Omega}{2} (J_z +a^\dagger a) t\right] \ . 
\end{equation}
In this frame, we study the quantum dynamics using the Hamiltonian $\hat{H}_{k}(t) = \hat{U}_{k}^\dagger(t)(\hat{H}(t)-i \frac{\partial}{\partial t})\hat{U}_{k}(t)$.
The explicit form of $\hat{H}_{k}(t)$ is 
\begin{eqnarray}
      \label{HamiltonianRotFrameSI}
	    \hat{H}_{\text{k}}(t)&=& \delta^{(k)} a^\dagger a + \left(\omega_0 \cos\left[\frac{2}{\sqrt{N}}\left(\frac{\Delta g \sin\Omega t}{\Omega}\right) \left(a^{\dagger}e^{\frac{ik \Omega t}{2}}+a e^{-\frac{ik \Omega t}{2}}\right)\right]-k \Omega/2\right) J_z
            \nonumber \\&&\nonumber
            +\frac{g}{\sqrt{N}}\left(a^{\dagger}e^{\frac{ik \Omega t}{2}}+a e^{-\frac{ik \Omega t}{2}}\right)\left(J_{+}e^{\frac{ik \Omega t}{2}}+J_{-}e^{\frac{-ik \Omega t}{2}}\right)
            \nonumber \\&&\nonumber
            -i\frac{\omega}{\sqrt{N}}\left(\frac{\Delta g \sin\Omega t}{\Omega}\right)\left(a^{\dagger}e^{\frac{ik \Omega t}{2}}-a e^{-\frac{ik \Omega t}{2}}\right)\left(J_{+}e^{\frac{ik \Omega t}{2}}+J_{-}e^{\frac{-ik \Omega t}{2}}\right)   
             \nonumber \\&& \nonumber
            +\frac{\omega}{N}\left[\left(\frac{\Delta g \sin\Omega t}{\Omega}\right)\left(J_{+}e^{\frac{ik \Omega t}{2}}+J_{-}e^{\frac{-ik \Omega t}{2}}\right)\right]^2
            \nonumber \\&&  
            -i\frac{\omega_0}{2} \sin\left[\frac{2}{\sqrt{N}}\left(\frac{\Delta g \sin\Omega t}{\Omega}\right) \left(a^{\dagger}e^{\frac{ik \Omega t}{2}}+a e^{-\frac{ik \Omega t}{2}}\right)\right] \left(J_{+}e^{\frac{ik \Omega t}{2}}-J_{-}e^{\frac{-ik \Omega t}{2}}\right),
\end{eqnarray}
where $\delta^{(k)}=\omega-k\Omega/2$. We then rewrite this equation using the operator identities 
\begin{align}
       \label{SinCosOperBesselSI} 
	\cos(\hat{\mathcal{O}}(t)\sin\Omega t) & =\mathcal{J}_0(\hat{\mathcal{O}}(t))+2\sum_{n=1}^{\infty}\mathcal{J}_{2n}(\hat{\mathcal{O}}(t))\cos(2n\Omega t) ,\nonumber \\
	\sin(\hat{\mathcal{O}}(t)\sin\Omega t)& = 2\sum_{n=0}^{\infty}\mathcal{J}_{2n+1}(\hat{\mathcal{O}}(t))\sin((2n+1)\Omega t) \ ,
\end{align}
where $\mathcal{J}_{m}(\cdotp)$ denotes the Bessel function of integer order $m$ \cite{AbramowitzSI}, and
the time-dependent operator argument is given by  $\hat{\mathcal{O}}(t)=\frac{2}{\sqrt{N}}\left(\frac{\Delta g }{\Omega}\right) \left(a^{\dagger}e^{\frac{ik \Omega t}{2}}+a e^{-\frac{ik \Omega t}{2}}\right)$.
Considering the power series expansion of the Bessel functions, it is possible to write the Hamiltonian Eq. \eqref{HamiltonianRotFrameSI} in the general form
\begin{equation}
      \label{HamiltonianFourierSI}
      \hat{H}_{\text{k}}(t)=\sum_{n=-\infty}^{\infty} \hat{h}^{(k)}_{n}\exp{(i n \Omega t)}  
      .
\end{equation}
In the rotating frame we are able to describe explicitly multiphoton processes that arise as a consequence of the interplay between the quantum cavity field and the classical external driving \cite{ShirleySI}. 

To perform the generalized RWA \cite{IrishSI,NoriSI,GrifoniSI}, we neglect the contributions due to virtual processes that ocurr faster than
the characteristic time scales of the system. Therefore, we consider 
only the the zero frequency component  $\hat{h}^{(k)}_{0}$ of the Hamiltonian Eq. \eqref{HamiltonianFourierSI}. For example, in the case of the $k=0$ resonance, a general condition of RWA validity in the strong driving regime should be satisfied for all $|n|>0$: $g,\ \omega_0 \lVert\mathcal{J}_{n}(\hat{\mathcal{O}}(0))\lVert, \ \omega \lVert\mathcal{J}_{n}(\hat{\mathcal{O}}(0))\rVert \ll \Omega,\Delta g$, where $\lVert\cdot\rVert$ is an operator norm. In contrast, in the weak driving regime $\Delta g/\Omega\ll 1$, we require $g,\ \omega ,\ \omega_0 \ll \Omega$. The norm of the Bessel function is always bounded, therefore we can summarize the conditions for weak and strong driving as $g,\ \omega ,\ \omega_0 \ll \Omega,\Omega\cdot\max(1,\Delta g/\Omega)$.

The full fundamental $k=0$ effective Hamiltonian drived with this approach reads:
\begin{eqnarray}
      \label{ZeroResonaceHamSI}
             h^{(0)}_{0} &=& \delta^{(0)}
            a^{\dagger}a +\delta^{(0)}_{0}\mathcal{J}_{0}\left(\frac{2\Delta g}{\Omega\sqrt{N}}(a^{\dagger}+a)\right)J_{z}
            \nonumber \\&& 
            +\frac{g}{\sqrt{N}}(a^{\dagger}+a)\left(J_{+}+J_{-}\right)
            + 2\omega\left (\frac{\Delta g}{\Omega}\right)^{2}\frac{J_{x}^{2}}{N}
            .
\end{eqnarray}
In the weak driving regime, the approximation to this Hamiltonian $\hat{h}^{(0)}_{0}$ up to second order in $\Delta g / \Omega$ is
\begin{eqnarray}
      \label{ZeroResonaceHamWeakDrivSI}
            h^{(0)}_{0} &\approx& \delta^{(0)}
            a^{\dagger}a +\delta^{(0)}_{0}J_{z}
            +\frac{g}{\sqrt{N}}(a^{\dagger}+a)\left(J_{+}+J_{-}\right)
            \nonumber \\&& 
            - \frac{\omega_{0}}{N}\left(\frac{\Delta g}{\Omega}\right)^2 (a^{\dagger}+a)^{2}J_{z}  + 2\omega\left (\frac{\Delta g}{\Omega}\right)^{2}\frac{J_{x}^{2}}{N}
            .
\end{eqnarray}
To second order, the $k=1$ effective Hamiltonian reads
\begin{eqnarray}
      \label{FirstResonaceHamSI}
            h^{(1)}_{0} & = & \delta^{(1)}
            a^{\dagger}a +\delta^{(1)}_{0}J_{z}
            \nonumber \\&& \nonumber  +\frac{g}{\sqrt{N}}\left(a^{\dagger}J_{-}+aJ_{+}\right)+\frac{\Delta g}{2\sqrt{N}}\left (a^{\dagger}J_{+}+aJ_{-}\right)
            \nonumber \\&& 
            -2\omega_{0}\left(\frac{\Delta g}{\Omega}\right)^2 a^{\dagger}a \frac{J_{z}}{N}+\omega\left(\frac{\Delta g}{\Omega} \right)^2 \frac{J_{-}J_{+}}{N} 
            ,
\end{eqnarray}
and similarly for the $k=2$ case
\begin{eqnarray}
      \label{SecondResonaceHamSI}
            h^{(2)}_{0} & = & \delta^{(2)}
            a^{\dagger}a +\delta^{(2)}_{0}J_{z} +\frac{g}{\sqrt{N}}\left(a^{\dagger}J_{-}+aJ_{+}\right)
            \nonumber \\&& \nonumber  -2\omega_{0}\left(\frac{\Delta g}{\Omega}\right)^2 a^{\dagger}a \frac{J_{z}}{N}+\omega\left(\frac{\Delta g}{\Omega} \right)^2 \frac{J_{-}J_{+}}{N} 
            \nonumber \\&& 
            + \frac{\omega_{0}}{2N}\left(\frac{\Delta g}{\Omega}\right)^2 ((a^{\dagger})^{2}+a^{2})J_{z} - \frac{\omega}{4N}\left(\frac{\Delta g}{\Omega}\right)^2 (J_{+}^{2}+J_{-}^{2})
            .
\end{eqnarray}

\section{The thermodynamic limit}
To address the expansion of the effective Hamiltonian $\hat{h}^{(0)}_{0}$
we begin by considering the Holstein-Primakoff representation of the angular momentum algebra \cite{HolsteinPrimakoffSI}
\begin{eqnarray}
        \label{HolsteinPrimakoffSI}
              J^{z} = b^\dagger b - N/2, \ J^{-}  = \sqrt{N - b^\dagger b} \: b, \ J^{+} = b^\dagger \sqrt{N - b^\dagger b} .             
\end{eqnarray}
In this representation, the driven Dicke Hamiltonian is written in terms of two bosonic modes $\hat{a}$ and $\hat{b}$.
We introduce 
macroscopic displacements $X\sqrt{N/2}$, $Y\sqrt{N/2}$ of order $\sqrt{N}$ for the bosonic $\hat{a}$ and for the atomic $\hat{b}$ modes, respectively. These macroscopic displacements are defined by
\begin{eqnarray}
        \label{DisplacementRotSI}
	      a = c\pm X\sqrt{\frac{N}{2}}, \qquad
	      b = d \mp Y\sqrt{\frac{N}{2}} .
\end{eqnarray}
In order to describe the QPT, we derive effective quadratic bosonic Hamiltonians for both the normal phase and the
superradiant phase \cite{BrandesSI}.
In the thermodynamic limit $N \rightarrow \infty$, we perform a series expansion of the Hamiltonian $\hat{h}^{(0)}_{0}$ in powers of $\sqrt{N/2}$
\begin{equation}
        \label{HamiltonExpSI}
	     \hat{h}^{(0)}_{0} \approx \hat{h}^{(0)}_{Q}(c,d,c^{\dagger},d^{\dagger}) + \sqrt{\frac{N}{2}}\ \hat{h}^{(0)}_{L}(c,d,c^{\dagger},d^{\dagger})+\frac{N}{2}\ E_G(X,Y) ,
\end{equation}
where $\hat{h}^{(0)}_{Q}$ is the desired effective quadratic bosonic Hamiltonian (depending on the choice of the macroscopic displacements Eq. \eqref{DisplacementRotSI}), $\hat{h}^{(0)}_{L}$ contains linear bosonic terms, and $E_G(X,Y)$ is the lowest quasienergy.

\end{widetext}


\begin{thebibliography}{03}
%
\bibitem{Dicke} R. H. Dicke, Phys. Rev. {\bf 93}, 99 (1954).
%
\bibitem{Hepp-Lieb} K. Hepp and E.H. Lieb, Phys. Rev. A {\bf 8}, 2517 (1973).
%
\bibitem{Brandes} C. Emary and T. Brandes, 
Phys. Rev. Lett. {\bf 90}, 044101 (2003);
Phys. Rev. E {\bf 67}, 066203 (2003).
%
\bibitem{Ciuti} P. Nataf and C. Ciuti, Nature Commun. {\bf 1}, 1 (2010).
%
\bibitem{Marquardt} O. Viehmann, J. von Delft, and F. Marquardt, Phys. Rev. Lett. {\bf 107}, 113602 (2011).
%
\bibitem{Esslinger} K. Baumann, C. Guerlin, F. Brennecke, and T. Esslinger, Nature (London) {\bf 464}, 1301 (2010).
%
\bibitem{Altland} A. Altland, V. Gurarie, T. Kriecherbauer, and A. Polkovnikov, Phys. Rev. A {\bf 79}, 042703 (2009).
%
\bibitem{Polkovnikov} A. Polkovnikov , K. Sengupta , A. Silva , and M. Vengalattore, Rev. Mod. Phys. {\bf 83}, 863 (2011).
%
\bibitem{Zurek} W. Zurek, U. Dorner, and P. Zoller, Phys. Rev. Lett. {\bf 95}, 105701 (2011).
%
\bibitem{Mottl} K. Baumann, R. Mottl, F. Brennecke, and T. Esslinger, Phys. Rev. Lett. {\bf 107}, 140402 (2011).
%
\bibitem{Hangui} J. Gong, L. Morales-Molina, and P. H\"anggi, Phys. Rev. Lett. {\bf 103}, 133002 (2009).
%
\bibitem{Holthaus} A. Eckardt, C. Weiss, and M. Holthaus, Phys. Rev. Lett. {\bf 95}, 260404 (2005).
%
\bibitem{Arimondo} H. Lignier, C. Sias, D. Ciampini, Y. Singh, A. Zenesini, O. Morsch, and E. Arimondo, Phys. Rev. Lett. {\bf 99}, 220403 (2007).
%
\bibitem{Gunter} G. G\"unter et al., Nature (London) {\bf 458}, 178 (2009).
%
\bibitem{DeLiberato} S. De Liberato, D. Gerace, I. Carusotto, and C. Ciuti, Phys. Rev. A 
{\bf 80}, 053810 (2009).
%
\bibitem{CarmichaelAlsing} P. Alsing, D.-S. Guo, and H. J. Carmichael, Phys. Rev. A {\bf 45}, 5135 (1992).
%
\bibitem{Peano} V. Peano and M. Thorwart, Phys. Rev. B {\bf 82}, 155129 (2010).
%
\bibitem{Grifoni} J. Hausinger and M. Grifoni, Phys. Rev. A {\bf 83}, 030301 (R) (2011).
%
\bibitem{Vedral} G. Vacanti, S. Pugnetti, N. Didier, M. Paternostro, G. M. Palma, R. Fazio, and V. Vedral, arXiv:1107.0178v1.
%
\bibitem{Creffield} C. E. Creffield, and T. S. Monteiro, Phys. Rev. Lett. {\bf 96}, 210403 (2006).
%
\bibitem{Lindner} N. H. Lindner, G. Refael, and V. Galitski, Nat. Phys. {\bf 7}, 490 (2011).
%
\bibitem{Tanaka} J. Inoue, and A. Tanaka, Phys. Rev. Lett. {\bf 105}, 017401 (2011).
%
\bibitem{Bastidas10} V. M. Bastidas, J. H. Reina, C. Emary, and T. Brandes, Phys. Rev. A {\bf 81}, 012316 (2010).
%
\bibitem{Arnold}
V. I. Arnold, {\it Mathematical Methods of Classical Mechanics}
(Springer-Verlag, New York, 1978).
%
\bibitem{Popov} A.M. Perelomov and V. S. Popov, Teor. Mat. Fiz. {\bf 1}, 360 (1969).
%
\bibitem{Tureci} B. \"{O}ztop, M. Bordyuh, \"{O}. E. M\"{u}stecapl{\i}o\u{g}lu, and H. E. T\"{u}reci, arXiv:1107.3108.
%
\bibitem{Shirley} J. H. Shirley, Phys. Rev. {\bf 138}, B979 (1965). 
%
\bibitem{Irish} E. K. Irish, Phys. Rev. Lett. {\bf 99}, 173601 (2007).
%
\bibitem{Nori} S. Ashhab, J. R. Johansson, A. M. Zagoskin, and F. Nori, Phys. Rev. A {\bf 75}, 063414 (2007).
%
\bibitem{SI} See supplementary material for additional details.
%
\bibitem{AparicioSimons} M. Aparicio Alcalde, and B. M. Pimentel, Physica A {\bf 390}, 3385 (2011); J. Keeling, M. J. Bhaseen, and B. D. Simons, Phys. Rev. Lett. {\bf 105}, 043001 (2010).
%
\bibitem{Footnote1} Note that this is different from previous proposals, e.g. D. Nagy, G. K\'onya, G. Szirmai, and P. Domokos, Phys. Rev. Lett. {\bf 104}, 130401 (2010), and F. Dimer, B. Estienne, A. S. Parkins, and H. J. Carmichael, Phys. Rev. A {\bf 75}, 013804 (2007), where the no-go theorem is overcome by considering effective degrees of freedom not bound by the sum rule.
%
\bibitem{1OPT} K. Binder, Rep. Prog. Phys. {\bf 50}, 783 (1987).
%
\end{thebibliography}

\begin{thebibliography}{03}
%
\bibitem{AbramowitzSI} M. Abramowitz and I. A. Stegun, \emph{Handbook of Mathematical Functions with Formulas, Graphs and Mathematical Tables}, edited by M. Abramowitz and I. A. Stegun (Dover, New York, 1972).
%
\bibitem{ShirleySI} J. H. Shirley, Phys. Rev. {\bf 138}, B979 (1965). 
%
\bibitem{IrishSI} E. K. Irish, Phys. Rev. Lett. {\bf 99}, 173601 (2007).
%
\bibitem{NoriSI} S. Ashhab, J. R. Johansson, A. M. Zagoskin, and F. Nori, Phys. Rev. A {\bf 75}, 063414 (2007).
%
\bibitem{GrifoniSI} J. Hausinger and M. Grifoni, Phys. Rev. A {\bf 83}, 030301 (R) (2011).
%
\bibitem{HolsteinPrimakoffSI} T. Holstein and H. Primakoff, Phys. Rev. {\bf 58}, 1098 (1949).
%
\bibitem{BrandesSI} C. Emary and T. Brandes, Phys. Rev. E {\bf 67}, 066203 (2003).
%
\end{thebibliography}
\end{document}